\documentclass[a4paper]{article}

\usepackage{INTERSPEECH2020}

\title{Parallel Rescoring with Transformer for Streaming On-Device Speech Recognition}
\name{Wei Li$^*$\thanks{$^*$Equal contribution.}, James Qin$^*$, Chung-Cheng Chiu, Ruoming Pang, Yanzhang He}
\address{Google Inc., USA}
\email{\{mweili, jamesqin, chungchengc, rpang, yanzhanghe\}@google.com}

\begin{document}

\maketitle
\begin{abstract}
Recent advances of end-to-end models have outperformed conventional models through employing a two-pass model. The two-pass model provides better speed-quality trade-offs for on-device speech recognition, where a $1st$-pass model generates hypotheses in a streaming fashion, and a $2nd$-pass model re-scores the hypotheses with full audio sequence context. The $2nd$-pass model plays a key role in the quality improvement of the end-to-end model to surpass the conventional model.
One main challenge of the two-pass model is the computation latency introduced by the $2nd$-pass model.
Specifically, the original design of the two-pass model uses LSTMs for the $2nd$-pass model, which are subject to long latency as they are constrained by the recurrent nature and have to run inference sequentially. In this work we explore replacing the LSTM layers in the $2nd$-pass rescorer with Transformer layers, which can process the entire hypothesis sequences \textit{in parallel} and can therefore utilize the on-device computation resources more efficiently. Compared with an LSTM-based baseline, our proposed Transformer rescorer achieves more than $50\%$ latency reduction with quality improvement.

\end{abstract}


\section{Introduction}

There has been a growing interest in building on-device streaming speech recognition models, which provide recognition results instantly as words are being spoken~\cite{rnnt}.  Such models make predictions based on partial context under strict latency requirements~\cite{streaming-e2e,pre-rescore,latency_mocha_msr}. As a result the streaming models tend to be less accurate than non-streaming models, which have access to the entire utterance.

Previous work have shown that this issue can be alleviated by combining a second-pass rescoring model \cite{two-pass} with streaming models, where the rescoring model uses the Listen, Attend, and Spell (LAS) architecture \cite{las}.  LAS has access to the full context of the utterance and therefore provides better quality than the streaming models~\cite{chiu2018state}.
From user's perspective, such a two-pass speech model exhibits the advantages of both streaming and non-streaming models---words are recognized as they are spoken and the final results have high accuracy.

The canonical architecture of the LSTM-based LAS model, however, is designed for beam search and is not efficient as a $2nd$-pass rescoring model.  The LSTM~\cite{lstm} layers process hypothesis tokens \textit{sequentially}, with temporal dependency between timesteps. 
On the other hand, for the $2nd$-pass rescoring, all hypothesis tokens are available. A more efficient design of the rescorer model will be to rescore all tokens in parallel.

In recent years there have been a growing success in applying Transformer~\cite{transformer} for machine translation and language modeling~\cite{t5}, and speech recognition~\cite{karita2019comparative, qian_tt, fb_tt, speech-xformer}.  Transformer applies self-attention to capture the sequential relation among input features, and therefore does not have the recurrent constraint. This allows Transformer to compute self-attention in parallel and significantly increase the computation efficiency. The Transformer architecture proposed in \cite{transformer} consists of an encoder and a decoder, where each decoder layer has an additional cross-attention that summarizes the encoder output based on the self-attention output.

In this work, we address the sequential dependency issue of the original LSTM-based rescoring model with Transformer. Specifically, the paper proposes to use Transformer as the second-pass rescorer for parallel rescoring of hypothesis tokens. Unlike beam search, where the Transformer decoder still has to run autoregressively, the rescoring scenario allows parallel processing of the full hypothesis sequence. Such parallelism reduces the lengths of temporal dependency paths from $\mathcal{O}(n)$ to $\mathcal{O}(1)$, where $n$ corresponds to the hypothesis length. This allows the Transformer rescorer to utilize on-device computation capacity much more efficiently. We further improve the inference speed of the Transformer rescorer by reducing the number of cross-attention in the decoder. The Transformer rescorer improves the Word Error Rate (WER) of Google’s voice search query test set to $5.7\%$ from $6.0\%$ with LSTM rescoring. On Librispeech~\cite{Panayotov2015} the Transformer rescorer improves the WER to $3.9\%$ on test clean and $9.8\%$ on test other compared to $4.0\%$ and $10.0\%$ with LSTM rescoring. The $90$th percentile second-pass latency, benchmarked on a Google Pixel4 phone on CPUs, is reduced to $57$ms from previous $127$ms with LSTM rescoring.
\begin{figure}[t]
  \centering
  \includegraphics[width=0.9\columnwidth]{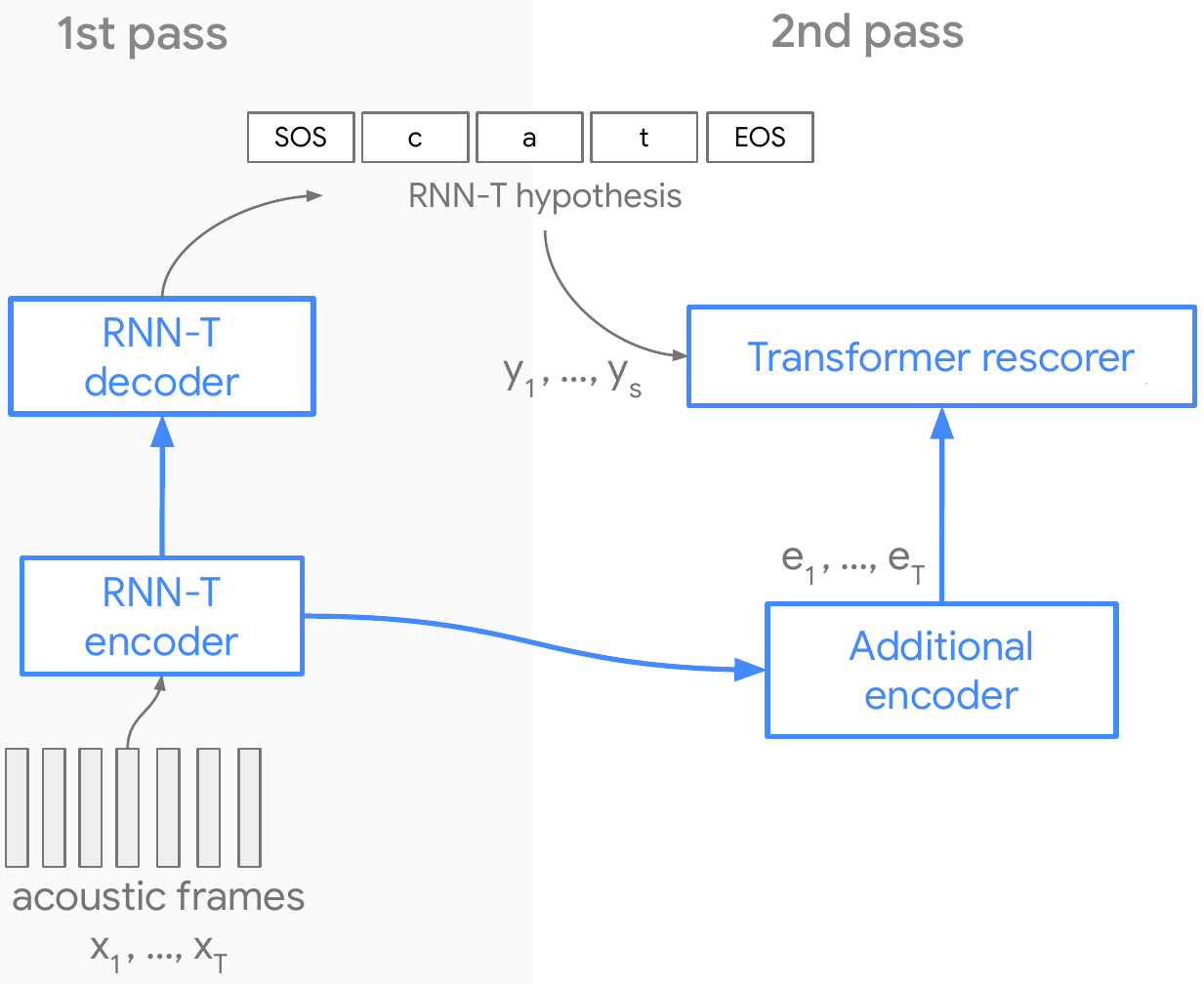}
  \caption{The architecture of two-pass model with Transformer.}
  \label{fig:rescorer_architecture}
\end{figure}

\section{Transformer Rescorer}

\subsection{Two-Pass Model}

A two-pass model consists of a $1st$-pass model and a $2nd$-pass model. Here we use RNN-T~\cite{Graves2012,Graves2013} as the $1st$-pass model and Transformer for the $2nd$-pass model. Specifically, our Transformer-based two-pass model, as demonstrated in Figure~\ref{fig:rescorer_architecture}, consists of four components: RNN-T encoder, RNN-T decoder, additional encoder, and Transformer decoder as the rescorer.  The input acoustic frames are denoted as $x = (x_1, ..., x_T)$, where $x_t \in R^d$ are stacked log-mel filterbank energies ($d = 512$) and $T$ is the number of frames in $x$. In the $1st$-pass, each acoustic frame $x_t$ is passed through RNN-T encoder, consisting of a multi-layer LSTM \cite{lstm}, to get encoder output. RNN-T decoder takes the acoustic features from RNN-T encoder to generate the hypotheses in a streaming fashion, denoted as $y = (y_1, . . ., y_s)$ where $s$ is the label sequence length.
Here $y$ is a sequence of word-piece tokens \cite{vocabulary}. In the $2nd$-pass, the full output of the RNN-T encoder is passed to a small additional encoder to generate $e_1,...,e_T$, which is then passed to Transformer decoder.  The additional encoder is added as it is found to be useful to adapt the encoder output to be more suitable for the second-pass model \cite{streaming-e2e}. The RNN-T model structure and the additional encoders are exactly the same as \cite{streaming-e2e}. During training, the Transformer decoder computes output label sequence according to the full audio sequence $e_1, ..., e_T$. More details about the rescorer training is elucidated in Section~\ref{sec:training}.  During decoding, the Transformer decoder rescores multiple top hypotheses from RNN-T, $y_1, ..., y_s$.  

\subsection{Transformer Rescorer Architecture}

\begin{figure}[t]
  \centering
  \includegraphics[width=0.8\columnwidth]{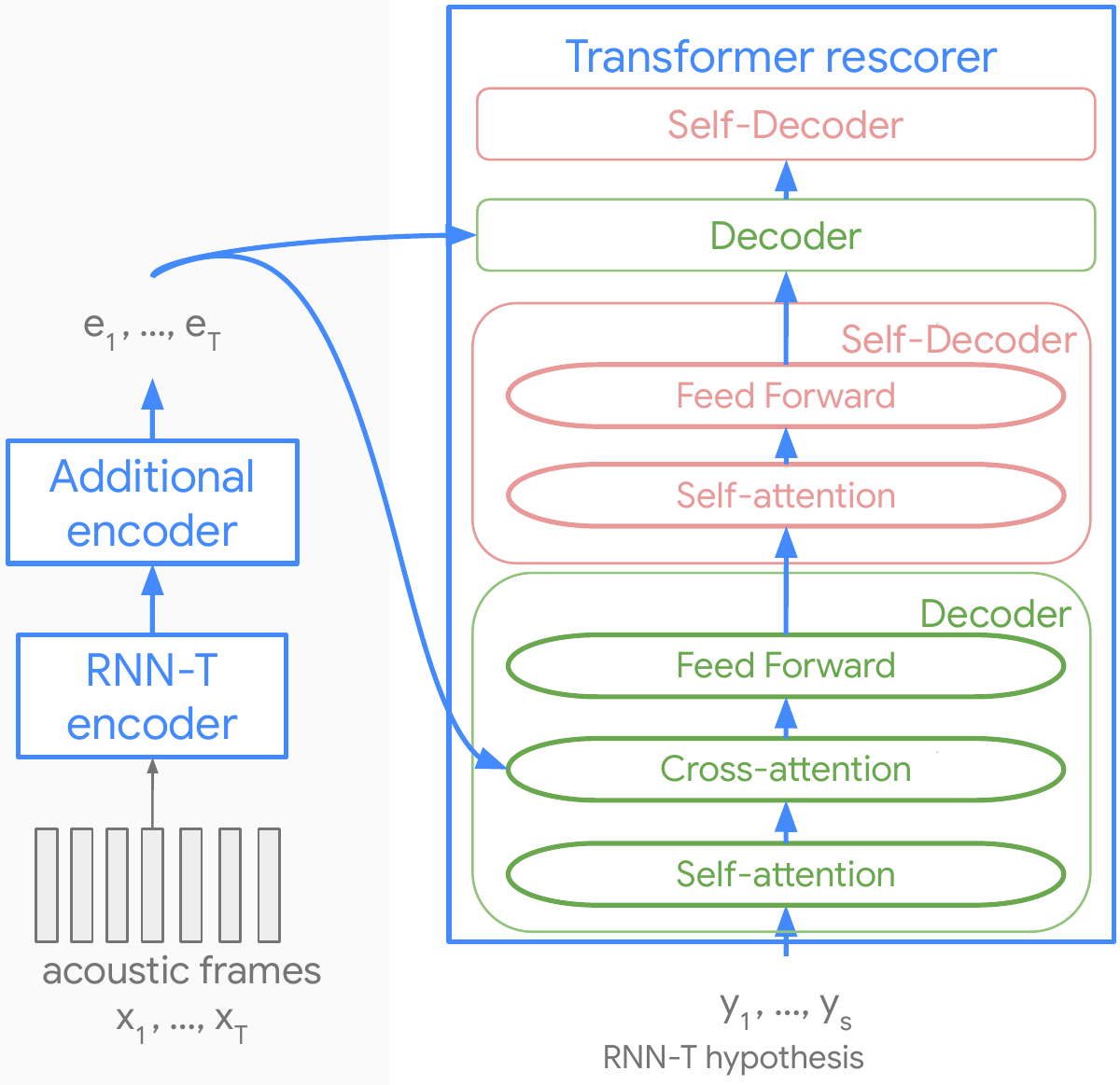}
  \caption{Transformer rescorer. The Transformer rescorer combines conventional Transformer decoders (containing cross-attention) and Transformer self-decoders (without cross-attention) for more efficient inference. The figure omits the normalization and residual links to simplify the illustration.}
  \label{fig:transformer}
\end{figure}

The architecture of our Transformer rescorer is based on the conventional Transformer decoder~\cite{transformer} with some cross-attention layers being removed. The conventional Transformer decoder layer contains both the self-attention and the cross-attention, where the query of the cross-attention originates from the output of the self-attention. In the Transformer rescorer, we improve the rescorer efficiency by removing the cross-attention from some decoder layers and interleave those layers with the conventional decoder layers. The decoder layer without the cross-attention shares the same architecture as the conventional Transformer encoder layer~\cite{transformer}. The architecture of the resulting rescorer is illustrated in Figure~\ref{fig:transformer}, where layers without cross-attention are annotated as \emph{self-decoder}. The Transformer rescorer takes the RNN-T's hypothesis as input and feed the tokens to the self-attention layer. And the cross-attention layers attend to the encoder output to summarize the acoustic signals. In our rescorer model, there are $4$ Transformer layers, each with the attention model dimension $d_{model}=640$ and feed forward dimension $d_{ff}=2560$. Both cross-attention and self-attention layers use multi-headed attention with $8$ heads. The rescorer model has $27.6M$ parameters.

Our design of keeping only two cross-attention layers in the rescorer is based on observing the attention mechanism of the Transformer decoder. In the first Transformer decoder layer, the self-attention conditions only on the hypothesis tokens, therefore the resulting cross-attention generates its query solely based on language modeling information. The missing of acoustic information on generating attention query inherently limit the effectiveness of the first cross-attention. After the first cross-attention layer, the output of the first decoder layer contains acoustic information, and the following decoder layers can condition on both the acoustic and language modeling information to generate effective cross-attention queries. Thus, it is critical to have the second cross-attention layer in the decoder. On the other hand, the additional cross-attention layers beyond the second one do not introduce additional modality and have diminishing returns in terms of the model quality. As a comparison, the cross-attention of the LAS model conditions on both the previous attention context and the text tokens, and requires only one cross-attention in the decoder. We demonstrated these property with an ablation study in Section~\ref{sec:exp}.

\subsection{Rescorer Training} \label{sec:training}

Same with the LAS rescoring training described in \cite{two-pass}, Transformer rescorer model is trained after the $1st$-pass model training.  During $2nd$-pass training, RNN-T encoder and RNN-T decoder are freezed.  Additional encoder and Transformer rescorer are trained in two stages: cross entropy (CE) and minimum word error rate (MWER) training \cite{mwer}. During CE training, frozen RNN-T encoder generates the acoustic features for additional encoder, and Transformer rescorer is trained to predict groundtruth sequence with the full audio context from additional encoder and the prefix of the label sequence context: $p(y_l|x, y_1...y_{l-1})$, where $l$ is the label to predict. During MWER training, the Transformer rescorer is trained to re-rank the hypotheses generated from RNN-T, which bridges the gap from CE training to inference \cite{two-pass}.  More specifically, given acoustic input $x$, groundtruth transcript $y\ast$, the probability computed by rescorer model $P(y_m|x)$ for any given target sequence $y_m$, and a set of hypotheses $H_m = {h_1, . . . , h_b}$ where b is the beam-size, the MWER loss is defined as
\begin{equation}
L_\text{MWER}(x, y\ast) = \sum_{y_m \in H_m(x)}P'(y_m|x, H_m)\left[W'(y_m, y\ast) - \widehat{W}\right] \nonumber
\end{equation}
where $P'(y_m|x, H_m) = \frac{P (y_m|x)}{\sum_{y_i \in H_m} P(y_i|x)}$ represents the conditional
probability the Transformer rescorer assigns to hypothesis $y_m$ among all hypotheses in $H_m$, and $W' (y\ast, y_m)$ is the number of word errors of $y_m$, and $\widehat{W}$ is the average number of word errors among $H_m$. In our MWER training we use the N-Best approximation approach for calculating the expected word errors \cite{mwer}.

\vspace{-0.1in}
\section{Quality Experiments}\label{sec:exp}
\subsection{Experiment Setup}

We conduct experiments on the Librispeech~\cite{Panayotov2015} dataset and a large-scale internal dataset. We use SpecAugment~\cite{park2019specaugment} with the same configuration as described in~\cite{largespecaugment} during training. Similar to~\cite{streaming-e2e}, we apply constant learning rate and maintain Exponential moving average (EMA) \cite{ema} of the weights during training, and use the EMA weights for evaluation. Both LSTM and Transformer rescorer are trained with CE and MWER. The N-Best size of MWER training is $4$, which matches the rescoring behavior during evaluation, where top $4$ hypotheses from RNN-T are used for rescoring. The prediction targets are $4096$ word pieces \cite{vocabulary} derived using a large corpus of text transcripts. The LSTM-based rescorer has size $33M$ and the Transformer has $27.6M$ parameters. All models are implemented in Tensorflow \cite{tf} using the Lingvo \cite{lingvo} toolkit and trained on $8 \times 8$ Tensor Processing Units (TPU) slices with a global batch size of $4096$.

\subsection{Librispeech Experiment}

In this experiment, the models are trained on the Librispeech 960h training set and evaluated on the clean and noisy test sets without an external language model. In order to maintain low-latency streaming speech recognition, the $1st$-pass RNN-T models in all the compared systems use a uni-directional LSTM encoder with 0 right context frame. As is shown in Table \ref{tab:libri_wer}, both the LSTM rescorer and the Transformer rescorer significantly improve the WER of the clean and noisy test sets compared to the RNN-T only model with $10$-$20\%$ relative improvement, alleviating the limited context problem for the $1st$-pass model while still maintaining low-latency streaming recognition. The Transformer rescorer further improves the WER slightly over the LSTM rescorer, and also significantly reduce the $2nd$-pass latency, which is studied in detail in Section~\ref{sec:latency_optimizations}.

\begin{table}[t]
  \caption{Librispeech test sets word error rate}
  \label{tab:libri_wer}
  \centering
  \begin{tabular}{c c c}
    \toprule
    \multicolumn{1}{c}{\textbf{Model}} & 
    \multicolumn{1}{c}{\textbf{Test clean}} & 
    \multicolumn{1}{c}{\textbf{Test other}}  \\
    \midrule
    RNN-T only & $4.9$ & $11.2$~~~              \\
    LSTM rescorer & $4.0$ & $10.0$~~~            \\
    Transformer rescorer & $3.9$ & $9.8$~~~     \\
    \bottomrule
  \end{tabular}
\end{table}

\subsection{Large Scale Experiment on Voice Search}
We perform a large scale experiment on an internal task, Google Voice Search, and show the proposed Transformer rescorer is also effective. In this experiment, the models are trained on a multi-domain training set as described in \cite{multidomain}. These multi-domain utterances span domains of search, farfield, telephony and YouTube. The test set includes $\sim14K$ Voice-search utterances (VS) extracted from Google traffic. All datasets are anonymized and hand-transcribed.  The transcription for YouTube utterances is done in a semi-supervised fashion \cite{semi1,semi2}. Following~\cite{li2020towards, chang2019joint, streaming-e2e}, we train the first-pass RNN-T to also emit the end-of-sentence decision to reduce the endpointing latency, allowing 2nd-pass rescoring to execute early.

As is shown in Table \ref{tab:vs_wer}, the Transformer rescorer improves the WER from $6.0$ to $5.7$ on the VS test set compared with the LSTM rescorer, both of which are trained with CE and MWER. Compared with $1st$-pass model, the Transformer rescorer achieves relative $10\%$ WER improvement.

\begin{table}[t]
  \caption{Voice Search test set word error rate}
  \label{tab:vs_wer}
  \centering
  \begin{tabular}{lr}
    \toprule
    \multicolumn{1}{c}{\textbf{Model}} & \multicolumn{1}{c}{\textbf{VS}} \\
    \midrule
    RNN-T only & $6.4$             \\
    LSTM rescorer & $6.0$             \\
    Transformer rescorer CE  & $5.9$               \\
    Transformer rescorer MWER  & $5.7$       \\
    \bottomrule
  \end{tabular}
\end{table}

\subsection{Full Context Rescoring}
The additional capability that the Transformer rescorer can bring is to utilize the full hypothesis when rescoring every target token. The original LSTM-based rescorer scores each target token conditioned only on the tokens before it.  Specifically, the LSTM rescorer learns a conditional probability $p(y_t|x, y_0,...,y_{t-1})$ for each prediction target $y_t$ where $y$ denotes hypothesis tokens from RNN-T and $x$ denotes acoustic features. A conventional Transformer decoder uses causal self-attention and also learns $p(y_t|x, y_0,...,y_{t-1})$. We explored extending the self-attention to access also the future label context and as a result learns to score target tokens with $p(y_t|x, y)$.
During CE training, using groundtruth sequence as the full context makes the training target trivial. Thus we randomly swap different proportions of the groundtruth tokens that fed to the self-attention layer with alternative tokens sampled within the word-piece vocabulary. Some sentinel tokens like \textit{SOS, EOS, UNKNOWN} and RNN-T's blank symbol are excluded to be used as random tokens.  The prediction targets are the original groundtruth sequence. During MWER training, the RNN-T hypothesis is used as the decoder input to match the inference scenario.  With this experiment, $15\%$ random proportion works out the best and achieves the same $5.7\%$ WER on the voice search task. Thus, we report results with causal self-attention for the experiments throughout the paper.

\section{Latency Optimizations} \label{sec:latency_optimizations}

In this section, we measure the additional latency introduced by the $2nd$-pass rescorer on a Google Pixel4 phone on CPUs. For efficient on-device execution, all models are converted to TensorFlow Lite format with post-training dynamic range quantization using the TensorFlow Lite Converter~\cite{tflite_quanitzation}. Matrix multiplication is operated in 8-bits with little accuracy loss. The benchmark suite consists of 89 utterances with voice action queries.
The LSTM rescorer latency baseline is fully optimized and is measured with lattice rescoring with batching described in \cite{streaming-e2e}. 

\subsection{Effect of Cross-Attention Layers}

We investigate the impact of the number of cross-attention layers on quality and latency. As shown in Table~\ref{tab:cross_atten}, we start with cross-attention on the $1st$ decoder layer and gradually add more. We observe a noticeable quality improvement at first, which later quickly diminishes. Specifically, with $2$ cross-attentions the rescorer achieves a $0.4$ WER improvement than $1$ cross-attention, but no further improvement is realized by adding more of it. In addition, when $2$ cross-attentions are used, we find that applying them on the $1st$ and $3rd$ layers improves WER by $0.15$ than on the $1st$ and $2nd$ layers. In the end, by selectively applying cross-attention, we achieved a $\sim 20ms$ latency reduction (Table~\ref{tab:latency}) and a $12.3\%$ ($4M$) parameter size reduction without quality compromise.

\begin{table}[th]
  \caption{Effect of cross-attention layers}
  \label{tab:cross_atten}
  \centering
  \begin{tabular}{l  r}
    \toprule
    \multicolumn{1}{c}{\textbf{Cross attention layers}} & 
    \multicolumn{1}{c}{\textbf{WER}} \\
    \midrule
    1st & $6.1$~~~             \\
    1st \& 2nd & $5.8$~~~             \\
    1st \& 3rd  & $5.7$~~~               \\
    All 4 layers  & $5.7$~~~       \\
    \bottomrule
  \end{tabular}
\end{table}
\vspace{-0.2in}

\subsection{Parallelism in Transformer Rescoring}

\begin{figure}[t]
  \centering
  \includegraphics[width=0.9\columnwidth]{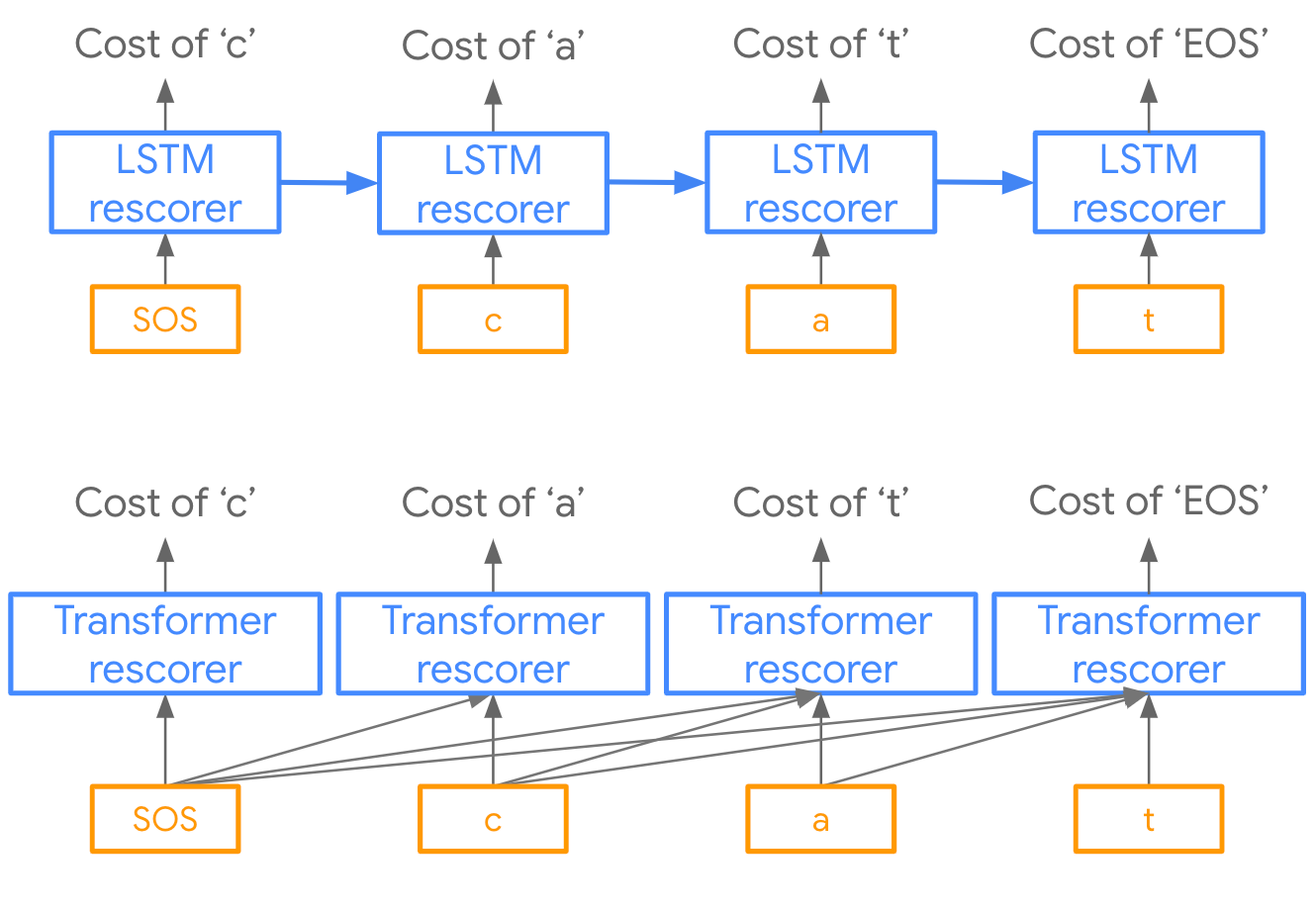}
  \caption{Parallel rescoring with Transformer.}
  \label{fig:parallel_rescoring}
\end{figure}

As is illustrated in Figure ~\ref{fig:parallel_rescoring}, with hypothesis labels ready from the $1st$-pass decoder output, Transformer rescorer can finish the computation in a single batch step as opposed to a series of sequential steps as in LSTM rescorer, which could better leverage multi-threading during inference. The batch size for transformer rescorer corresponds to
\begin{equation}
     \mbox{number\_of\_hyps} \times \mbox{hyp\_length} \times  \mbox{number\_of\_attention\_heads}. \nonumber
\end{equation}
Taking the utterance at the $90$th percentile latency as an example, with the top $4$ hypotheses used, the batch size is $4 \times 12 \times 8 = 384$. This large batch size provides better parallelism and as a result benefits more from using $2$ threads which reduces $35ms$ latency (Table~\ref{tab:latency}).  The multi-threading benefit is not witnessed in the LSTM-based rescorer. Potentially it might be due to (1) limited parallelism in LSTM, where batching is done within each inference step with a relatively smaller batch size being $\mbox{number\_of\_hyps} \times \mbox{number\_of\_gates}$ and (2) utilizing multi-threading within each inference step could introduce extra overhead due to context switch across inference steps and layers.

\subsection{Latency Measurements and Distributions}

An overall breakdown for latency optimizations is shown in Table~\ref{tab:latency}. The Transformer rescorer achieves a $55\%$ latency reduction compared to the LSTM rescorer, measured on the utterance with the $90$th percentile latency with the LSTM rescorer, which has $6s$ audio and $12$ word-piece tokens in the transcript.

The initial latency of the Transformer rescorer with $4$ cross-attention layer is $106ms$, which then improves to $92ms$ by keeping only $2$ cross-attentions.  Compared to $127ms$ from LSTM baseline, the $27\%$ latency improvement is from the reduced FLOPs.  Transformer rescorer with $4$ and $2$ cross-attentions provide a $15\%$ ($340M$) and $20\%$ ($320M$) FLOPs reduction compared to LSTM ($400M$).

Using two threads reduces the latency by an additional $35ms$ for Transformer rescorer, while the LSTM rescorer does not benefit from multi-threading.

\begin{table}[th]
  \caption{Computational latency for the Transformer rescorer with various optimizations, benchmarked on Pixel4 CPUs.}
  \label{tab:latency}
  \centering
  \begin{tabular}{l  r}
    \toprule
    \multicolumn{1}{c}{\textbf{Optimizations}} & 
    \multicolumn{1}{c}{\textbf{Latency(ms)}} \\
    \midrule
    Initial latency ($4$ cross attention) & $106$~~~               \\
    $2$ cross attention  & $92$~~~       \\
    Parallelism in two threads  & $57$~~~              \\
    \hline
    LSTM baseline & $127$~~~             \\
    \bottomrule
  \end{tabular}
\end{table}

\begin{figure}[t]
  \centering
  \includegraphics[width=0.9\columnwidth,height=2.4in]{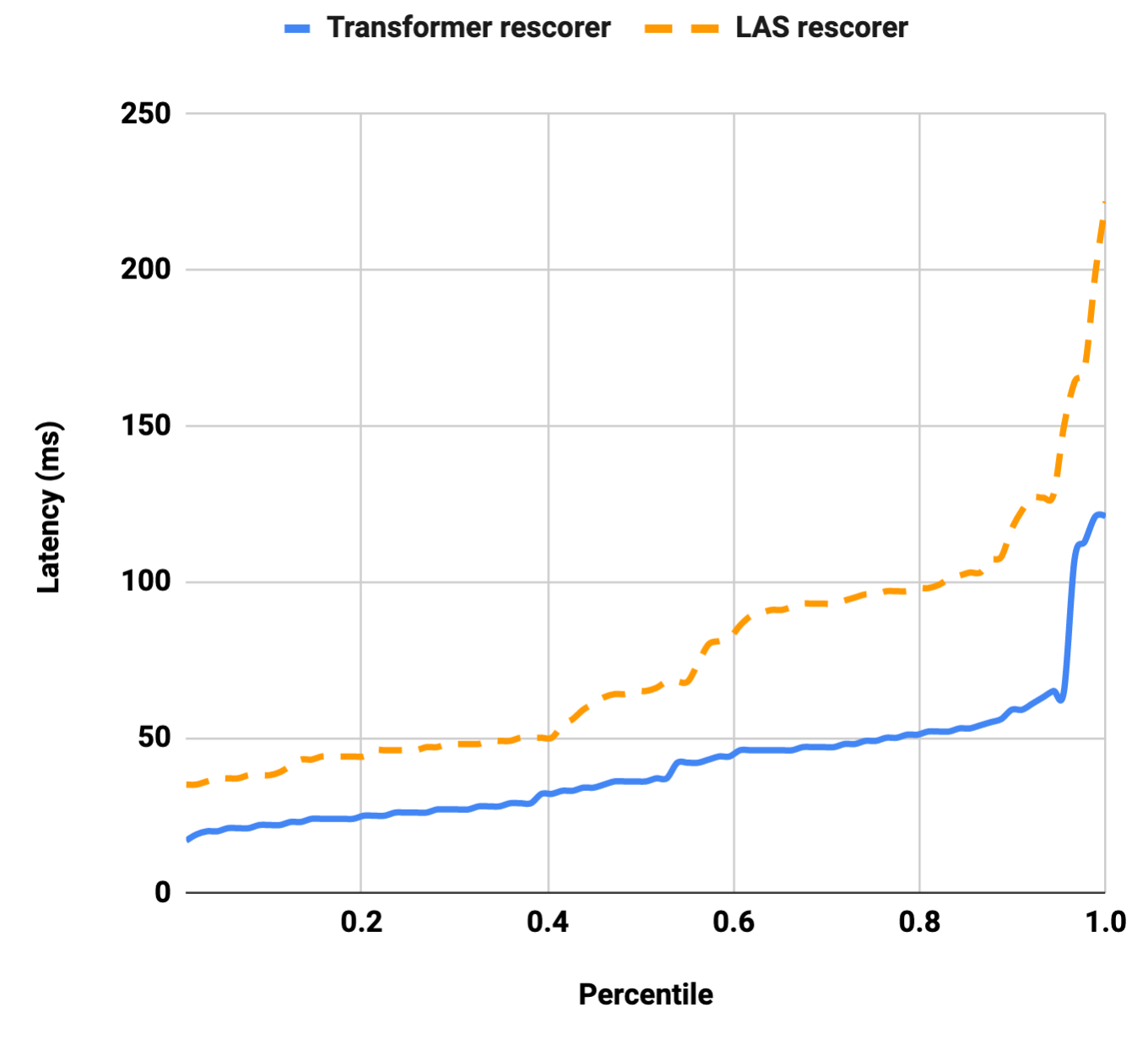}
  \caption{Latency comparison by percentile.}
  \label{fig:latency_comparison}
\end{figure}

We also compared the latency distribution over the full benchmark suite, demonstrated in Figure~\ref{fig:latency_comparison}. The speech time ranges from $1.5s$ to $9.3s$ in the benchmark. The output label sequence length varies from $3$ to $29$. Transformer rescorer is consistently $\sim 50\%$ faster than LSTM rescorer at almost every latency percentile.

\section{Conclusion}

In this work we present a Transformer rescorer for a two-pass model. Our proposed Transformer rescorer reduces more than $50\%$ of the on-device computation latency in second-pass model by taking advantage of the parallelism in Transformer decoder and reducing the number of cross attention layers. On a Google Voice Search task the Transformer rescorer achieves $5.7\%$ WER compared with $6.0\%$ of an LSTM rescorer. On Librispeech the Transformer rescorer achieves $3.9\%$ and $9.8\%$ WER on test clean and test other, also lower than $4.0\%$ and $10.0\%$ of the LSTM rescorer, respectively.

\section{Acknowledgements}
We thank TF-Lite team for the help to get Transformer model running on device, especially T.J. Alumbaugh, Jared Duke, Jian Li, Feng Liu and Renjie Liu. We are also grateful for the insightful discussions with Shuo-yiin Chang, Ian McGraw, Tara Sainath and Yonghui Wu. 

\bibliographystyle{IEEEtran}

\bibliography{mybib}

\begin{thebibliography}{10}
\providecommand{\url}[1]{#1}
\csname url@samestyle\endcsname
\providecommand{\newblock}{\relax}
\providecommand{\bibinfo}[2]{#2}
\providecommand{\BIBentrySTDinterwordspacing}{\spaceskip=0pt\relax}
\providecommand{\BIBentryALTinterwordstretchfactor}{4}
\providecommand{\BIBentryALTinterwordspacing}{\spaceskip=\fontdimen2\font plus
\BIBentryALTinterwordstretchfactor\fontdimen3\font minus
  \fontdimen4\font\relax}
\providecommand{\BIBforeignlanguage}[2]{{%
\expandafter\ifx\csname l@#1\endcsname\relax
\typeout{** WARNING: IEEEtran.bst: No hyphenation pattern has been}%
\typeout{** loaded for the language `#1'. Using the pattern for}%
\typeout{** the default language instead.}%
\else
\language=\csname l@#1\endcsname
\fi
#2}}
\providecommand{\BIBdecl}{\relax}
\BIBdecl

\bibitem{rnnt}
Y.~He, T.~N. Sainath, R.~Prabhavalkar, I.~McGraw, R.~Alvarez, D.~Zhao,
  D.~Rybach, A.~Kannan, Y.~Wu, R.~Pang, Q.~Liang, D.~Bhatia, Y.~Shangguan,
  B.~Li, G.~Pundak, K.~C. Sim, T.~Bagby, S.~yiin Chang, K.~Rao, and
  A.~Gruenstein, ``Streaming end-to-end speech recognition for mobile
  devices,'' in \emph{ICASSP 2019 - 2019 IEEE International Conference on
  Acoustics, Speech and Signal Processing (ICASSP)}, 2019.

\bibitem{streaming-e2e}
T.~N. Sainath, Y.~He, B.~Li, A.~Narayanan, R.~Pang, A.~Bruguier, S.-y. Chang,
  W.~Li, R.~Alvarez, Z.~Chen, and et~al., ``A streaming on-device end-to-end
  model surpassing server-side conventional model quality and latency,''
  \emph{ICASSP 2020 - 2020 IEEE International Conference on Acoustics, Speech
  and Signal Processing (ICASSP)}, May 2020.

\bibitem{pre-rescore}
S.-Y. Chang, B.~Li, D.~Rybach, Y.~He, W.~Li, T.~Sainath, and T.~Strohman,
  ``{{Low Latency Speech Recognition using End-to-End Prefetching}},'' in
  \emph{Proc. of Interspeech}, 2020.

\bibitem{latency_mocha_msr}
H.~Inaguma, Y.~Gaur, L.~Lu, J.~Li, and Y.~Gong, ``Minimum latency training
  strategies for streaming sequence-to-sequence asr,'' in \emph{ICASSP
  2020-2020 IEEE International Conference on Acoustics, Speech and Signal
  Processing (ICASSP)}.\hskip 1em plus 0.5em minus 0.4em\relax IEEE, 2020, pp.
  6064--6068.

\bibitem{two-pass}
T.~N. Sainath, R.~Pang, D.~Rybach, Y.~He, R.~Prabhavalkar, W.~Li, M.~Visontai,
  Q.~Liang, T.~Strohman, Y.~Wu, I.~McGraw, and C.-C. Chiu, ``{{Two-Pass
  End-to-End Speech Recognition}},'' in \emph{Proc. of Interspeech}, 2019.

\bibitem{las}
W.~Chan, N.~Jaitly, Q.~V. Le, and O.~Vinyals, ``Listen, attend and spell,''
  2015.

\bibitem{chiu2018state}
C.-C. Chiu, T.~N. Sainath, Y.~Wu, R.~Prabhavalkar, P.~Nguyen, Z.~Chen,
  A.~Kannan, R.~J. Weiss, K.~Rao, E.~Gonina, N.~Jaitly, B.~Li, J.~Chorowski,
  and M.~Bacchiani, ``State-of-the-art speech recognition with
  sequence-to-sequence models,'' in \emph{Proc. of ICASSP}, 2018.

\bibitem{lstm}
S.~Hochreiter and J.~Schmidhuber, ``Long short-term memory,'' \emph{Neural
  Comput.}, p. 1735–1780, Nov. 1997.

\bibitem{transformer}
A.~Vaswani, N.~Shazeer, N.~Parmar, J.~Uszkoreit, L.~Jones, A.~N. Gomez, L.~u.
  Kaiser, and I.~Polosukhin, ``Attention is all you need,'' in \emph{Advances
  in Neural Information Processing Systems 30}, 2017, pp. 5998--6008.

\bibitem{t5}
C.~Raffel, N.~Shazeer, A.~Roberts, K.~Lee, S.~Narang, M.~Matena, Y.~Zhou,
  W.~Li, and P.~J. Liu, ``Exploring the limits of transfer learning with a
  unified text-to-text transformer,'' \emph{JMLR}, 2019.

\bibitem{karita2019comparative}
S.~Karita, N.~Chen, T.~Hayashi, T.~Hori, H.~Inaguma, Z.~Jiang, M.~Someki,
  N.~E.~Y. Soplin, R.~Yamamoto, X.~Wang \emph{et~al.}, ``A comparative study on
  transformer vs rnn in speech applications,'' \emph{arXiv preprint
  arXiv:1909.06317}, 2019.

\bibitem{qian_tt}
Q.~Zhang, H.~Lu, H.~Sak, A.~Tripathi, E.~McDermott, S.~Koo, and S.~Kumar,
  ``Transformer transducer: A streamable speech recognition model with
  transformer encoders and rnn-t loss,'' \emph{ICASSP 2020 - 2020 IEEE
  International Conference on Acoustics, Speech and Signal Processing
  (ICASSP)}, May 2020.

\bibitem{fb_tt}
C.-F. Yeh, J.~Mahadeokar, K.~Kalgaonkar, Y.~Wang, D.~Le, M.~Jain, K.~Schubert,
  C.~Fuegen, and M.~L. Seltzer, ``Transformer-transducer: End-to-end speech
  recognition with self-attention,'' 2019.

\bibitem{speech-xformer}
L.~{Dong}, S.~{Xu}, and B.~{Xu}, ``Speech-transformer: A no-recurrence
  sequence-to-sequence model for speech recognition,'' in \emph{2018 IEEE
  International Conference on Acoustics, Speech and Signal Processing
  (ICASSP)}, 2018, pp. 5884--5888.

\bibitem{Panayotov2015}
V.~{Panayotov}, G.~{Chen}, D.~{Povey}, and S.~{Khudanpur}, ``Librispeech: An
  asr corpus based on public domain audio books,'' in \emph{Proc. of ICASSP},
  2015, pp. 5206--5210.

\bibitem{Graves2012}
A.~Graves, ``Sequence transduction with recurrent neural networks,''
  \emph{CoRR}, vol. abs/1211.3711, 2012.

\bibitem{Graves2013}
A.~Graves, A.~r.~Mohamed, and G.~Hinton, ``Speech recognition with deep
  recurrent neural networks,'' in \emph{Proc. of ICASSP}, 2013.

\bibitem{vocabulary}
M.~{Schuster} and K.~{Nakajima}, ``{Japanese and Korean Voice Search},'' in
  \emph{Proc. of ICASSP}, 2012, pp. 5149--5152.

\bibitem{mwer}
R.~Prabhavalkar, T.~N. Sainath, Y.~Wu, P.~Nguyen, Z.~Chen, C.-C. Chiu, and
  A.~Kannan, ``Minimum word error rate training for attention-based
  sequence-to-sequence models,'' \emph{2018 IEEE International Conference on
  Acoustics, Speech and Signal Processing (ICASSP)}, Apr 2018.

\bibitem{park2019specaugment}
D.~S. Park, W.~Chan, Y.~Zhang, C.-C. Chiu, B.~Zoph, E.~D. Cubuk, and Q.~V. Le,
  ``Specaugment: A simple data augmentation method for automatic speech
  recognition,'' in \emph{Interspeech}, 2019.

\bibitem{largespecaugment}
D.~S. Park, Y.~Zhang, C.-C. Chiu, Y.~Chen, B.~Li, W.~Chan, Q.~V. Le, and Y.~Wu,
  ``Specaugment on large scale datasets,'' in \emph{ICASSP}, 2020.

\bibitem{ema}
B.~Polyak and A.~Juditsky, ``{Acceleration of Stochastic Approximation by
  Averaging},'' \emph{SIAM Journal on Control and Optimization}, vol.~30,
  no.~4, 1992.

\bibitem{tf}
M.~Abadi, P.~Barham, J.~Chen, Z.~Chen, A.~Davis, J.~Dean, M.~Devin,
  S.~Ghemawat, G.~Irving, M.~Isard \emph{et~al.}, ``{Tensorflow: A System for
  Large-scale Machine Learning},'' pp. 265--283, 2016.

\bibitem{lingvo}
J.~Shen, P.~Nguyen, Y.~Wu, Z.~Chen, and et~al., ``Lingvo: a modular and
  scalable framework for sequence-to-sequence modeling,'' 2019.

\bibitem{multidomain}
A.~Narayanan, R.~Prabhavalkar, C.-C. Chiu, D.~Rybach, T.~Sainath, and
  T.~Strohman, ``{Recognizing Long-Form Speech Using Streaming End-to-End
  Models},'' in \emph{Proc. ASRU}, 2019.

\bibitem{semi1}
H.~Liao, E.~McDermott, and A.~Senior, ``Large scale deep neural network
  acoustic modeling with semi-supervised training data for youtube video
  transcription,'' in \emph{2013 IEEE Workshop on Automatic Speech Recognition
  and Understanding}, 2013.

\bibitem{semi2}
H.~Soltau, H.~Liao, and H.~Sak, ``{Neural Speech Recognizer: Acoustic-to-Word
  {LSTM} Model for Large Vocabulary Speech Recognition},'' in \emph{Proc. of
  Interspeech}, 2017, pp. 3707--3711.

\bibitem{li2020towards}
B.~Li, S.-Y. Chang, T.~N. Sainath, R.~Pang, Y.~He, T.~Strohman, and Y.~Wu,
  ``Towards fast and accurate streaming end-to-end asr,'' in \emph{ICASSP
  2020-2020 IEEE International Conference on Acoustics, Speech and Signal
  Processing (ICASSP)}.\hskip 1em plus 0.5em minus 0.4em\relax IEEE, 2020, pp.
  6069--6073.

\bibitem{chang2019joint}
S.-Y. Chang, R.~Prabhavalkar, Y.~He, T.~N. Sainath, and G.~Simko, ``Joint
  endpointing and decoding with end-to-end models,'' in \emph{ICASSP 2019-2019
  IEEE International Conference on Acoustics, Speech and Signal Processing
  (ICASSP)}.\hskip 1em plus 0.5em minus 0.4em\relax IEEE, 2019, pp. 5626--5630.

\bibitem{tflite_quanitzation}
\BIBentryALTinterwordspacing
``Tensorflow lite: Post-training quantization.'' [Online]. Available:
  \url{https://www.tensorflow.org/lite/performance/post\_training\_quantization}
\BIBentrySTDinterwordspacing

\end{thebibliography}

\end{document}